# VARIABILITY OF THE SPIN PERIOD OF THE WHITE DWARF IN THE MAGNETIC CATACLYSMIC BINARY SYSTEM EX HYA

Ivan L. Andronov, Vitalii V. Breus

Department «High and Applied mathematics»,
Odessa National Maritime University, Odessa, Ukraine;
e-mail: *ilandronov @ gmail.com, bvv_2004 @ ua.fm*

## Abstract

The observations of the two-periodic magnetic cataclysmic system EX Hya have been carried out, using the telescopes RC16 and TOA-150 of the Tzec Maun observatory. 6 nights of observations were obtained in 2010-2011 (alternatively changing filters VR). Also the databases of WASP, ASAS and AAVSO have been analyzed. Processing time series was carried out using the program MCV. We analyzed changes in the rotation period of the white dwarf, and based on our own and previously published moments of maximum. The ephemeris was determined for the maxima of the radiation flux associated with the rotation of the magnetic white dwarf: $T_{max}=2437699.89079(59) +0.0465464808(69) \cdot E - 6.3(2) \cdot 10^{-13} E^2$, which corresponds to the characteristic timescale of the rotation spin-up of $4.67(14) \cdot 10^6$ years. This contradicts the estimated value of the mass of the white dwarf of $0.42 M_\odot$, based on X-ray observations made by Yuasa et al (2010), however, is consistent with estimates of the masses of $0.79\ M_\odot$ (white dwarf) and $0.108\ M_\odot$ (red dwarf) previously published by Beuermann and Reinsch (2008), and the assumption that the capture of accreted plasma by magnetic field of the white dwarf is near the border of the Roche lobe. Analyzed moments do not support the assumption of Mauche et al (2009) for a statistically significant cubic term in the ephemeris. Despite the presence of outbursts in EX Hya, there are significant differences from the DO Dra, which supports the introduction to a detailed classification of the intermediate polars the groups of "outbursting intermediate polars" and "magnetic dwarf novae."

## Introduction

Intermediate polars, often called DQ Her star, are interacting binary systems with strong magnetic fields [1-3]. Gravity of the white dwarf leads to the gravitational capture of the part of the substance of the secondary component near the inner Lagrange points. Due to the Coriolis force, plasma flux deviates from the center line and forms an accretion disk around a white dwarf. A strong magnetic field destroys the inner part of the disk and leads to the formation of two accretion columns, which are one of the brightest sources of radiation in a wide spectral range from x - ray to infrared. The cyclotron radiation is characterized by the presence of polarization. The matter forms a shock wave heats up and settles on the surface of the white dwarf. Hot gas emits «bremsstrahlung» radiation in the hard (up to 50 Kev) X-ray range. Rare outbursts are possible (e.g. DO Dra [4]). Usually intermediate polars were classified as nova-like stars with relatively small changes in average per night light (up to 1-2 magnitude).

EX Hya was firstly noted as a variable star in the catalogue of Brun and Petit in 1957 [5]. Later it was continued to be classified as a dwarf nova (e.g. [6]). Papaloijzuu and Pringle [7] in 1980 noted that the ratio of photometric periods is close to the 2:3, Warner and Mac-Grau [8] discussed two models - modulation of the mass flow and the magnetic white dwarf (intermediate polar). Then, the object was classified either as dwarf nova, or as an intermediate polar [9], as the system displays characteristics of both classes of cataclysmic variables. We can talk about a group of «magnetic dwarf novae» or «outbursting intermediate polars» [4].

In this paper, we study variations of the rotation (spin) period of the white dwarf in the EX Hydra system. This work is part of the international project «Inter-Longitude Astronomy» [10] and the national project «Ukrainian Virtual Observatory» [11].

**Observations**

Observations were carried out using remotely-controlled telescopes TOA150 (15cm) and BigMak (35cm) at the Tzec Maun observatory (http://tzecmaun.org/). We obtained 5 nights of observations in 2010 and 1 night in 2011, 129 expositions in the V filter and 118 in the filter $R_c$. For processing of CCD frames, we used the program MuniWin (http://c-munipack.sourceforge.net/, http://munipack.astronomy.cz/). To improve the accuracy of measurements of brightness, we used the technique of "multiple comparison stars", described in [12, 13] and implemented in the computer program MCV (http://uavso.pochta.ru/mcv).

We obtained the tables of measurements of object and comparison stars. Journal of observations is presented in the Table 1. Comparison stars are marked in the Fig.1, a list of the characteristics defined by A. Henden (ftp://ftp.aavso.org/public/calib/exhya.dat), are shown in table 2. As the primary calibration star, we used C1. It's magnitude was assumed to be $Rc=11.854^m$. Unfortunately, for other comparison stars, the measurements in this photometric system are missing. The definition of coefficients of relation between the standard and instrumental photometric systems is impossible at this stage. Although we observed in two filters, and light curves were plotted in the instrumental systems V, Rc, bringing the color indices to the standard system it is impossible at this stage. However, our observations could be used in further works after calibration.

TABLE 1. Journal of observations.

| HJD start | HJD end | Mean brightness | Exposures | Filter |
|---|---|---|---|---|
| 55240.2869 | 55240.2984 | 13.448 ± 0.054 | 5 | V |
| 55240.2884 | 55240.2970 | 13.423 ± 0.041 | 4 | R |
| 55242.0073 | 55242.0370 | 13.400 ± 0.029 | 13 | V |
| 55242.0085 | 55242.0383 | 13.508 ± 0.023 | 12 | R |
| 55245.9186 | 55245.9617 | 13.421 ± 0.055 | 20 | V |
| 55245.9195 | 55245.9607 | 13.510 ± 0.056 | 18 | R |
| 55266.1399 | 55266.2316 | 13.368 ± 0.025 | 27 | V |
| 55266.1416 | 55266.2330 | 13.461 ± 0.026 | 25 | R |
| 55278.9391 | 55279.0253 | 13.421 ± 0.024 | 30 | V |
| 55278.9406 | 55279.0238 | 13.447 ± 0.023 | 29 | R |
| 55719.9281 | 55720.1044 | 13.189 ± 0.031 | 34 | V |
| 55719.9436 | 55720.1056 | 12.975 ± 0.035 | 30 | R |

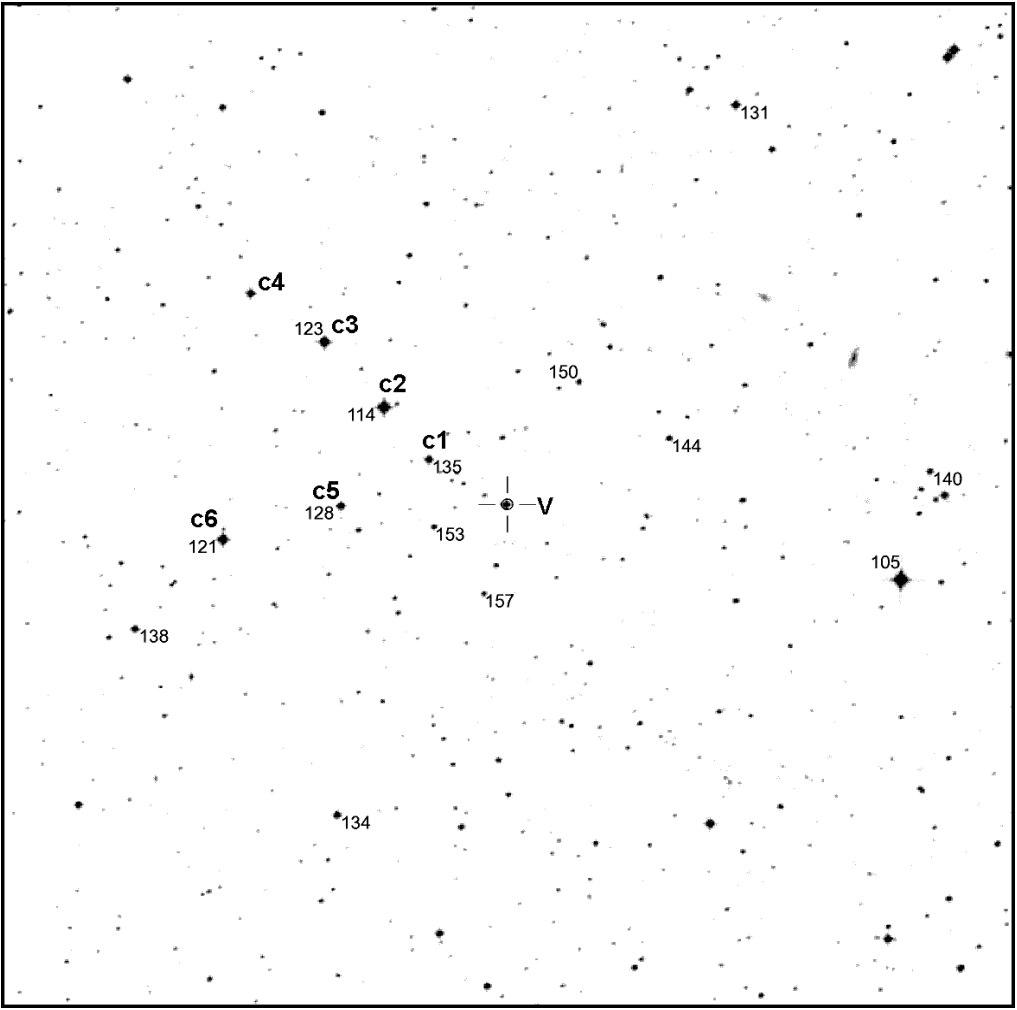

Fig. 1. Comparison stars in the field of EX Hya (we used C1-C6; that recommended by AAVSO are marked by numbers, e.g. 135 corresponds to V=13.5$^m$ rounded to 0.1$^m$). Coordinates of the center of the field R.A.=12$^h$52$^m$24.47$^s$, Dec. = -29$^d$14$^m$57.5$^s$ (J2000), the letter "v" in the centre indicates the variable EX Hya. Field size is 20'x20'

TABLE 2. Information about the comparison stars for EX Hya.

| AAVSO | ID | RA | Dec | V | $R_c$ |
|---|---|---|---|---|---|
| 105 | - | 12:51:48.59 | -29:16:27.3 | 10.477(043) | 9.731(101) |
| 114 | C2 | 12:52:35.39 | -29:12:59.4 | 11.428(59) | - |
| 121 | C6 | 12:52:50.08 | -29:15:39.1 | 12.122(56) | - |
| 123 | C3 | 12:52:40.86 | -29:11:41.2 | 12.253(60) | - |
| 128 | C5 | 12:52:39.42 | -29:14:58.4 | 12.775(56) | - |
| 131 | - | 12:52:03.59 | -29:06:57.3 | 13.078(33) | 12.521(43) |
| 134 | - | 12:52:39.71 | -29:21:09.8 | 13.375(23) | 12.812(55) |
| 135 | C1 | 12:52:31.39 | -29:14:02.2 | 13.542(60) | - |
| 138 | - | 12:52:57.96 | -29:17:26.1 | 13.788(00) | 13.271(39) |
| 140 | - | 12:51:45.97 | -29:14:16.1 | 14.042(23) | 13.506(69) |
| 144 | - | 12:52:09.60 | -29:13:36.8 | 14.385(60) | - |
| 150 | - | 12:52:17.81 | -29:12:29.2 | 14.953(55) | - |
| 153 | - | 12:52:30.94 | -29:15:24.1 | 15.291(52) | - |
| 157 | - | 12:52:26.41 | -29:16:44.1 | 15.694(56) | - |

**Photometric observations**

In the journal of observations is given for all nights. An example of the light curve obtained for one of the nights is shown in Fig.2. We choose 2-periodic model of variability for smoothing:

$$m(t)=m_0-r_1\cos(\omega_1(t-T_{01})) -r_2\cos(\omega_2(t-T_{02})) \qquad (1)$$

where $m(t)$ - is the smoothed value of brightness at time $t$, $m_0$ - average brightness of theoretical function (generally different from the sample mean, see, e.g. reviews [14,15]), $\omega_j=2\pi/P_j$, $r_j$ - semi-amplitude, $T_{0j}$ is the epoch for maxima of brightness (magnitude minima) of photometric wave with number $j$ and period $P_j$. Similar approximation was used during the study of other intermediate polars, e.g. BG CMi [16], MU Cam [17]. We discovered [17] the phenomenon of variability of phase of maximum of photometric wave associated with the rotation of the white dwarf, on orbital phase in MU Cam. From a theoretical point of view, this could be interpreted by modulation of the accretion flow structure by the magnetic field of a rapidly rotating white dwarf. From the observational point of view, this indicates a necessity of using (for the study of slow changes of phase) «mean maxima» (averaged values over the orbital period). Thus, the formula (1) is optimal for approximation of observations of intermediate polars, although for some of them (e.g. V405 Aur), it is necessary to take in account harmonics of the first and/or second periods.

As we took short intervals of observations for smoothing, we used published earlier values of the periods $P_1=P_{orb}=0.068233846^d$ (orbital period), $P_2=P_{rot}=0.046546504^d$ (rotation period of the white dwarf). Below we examine the period variations.

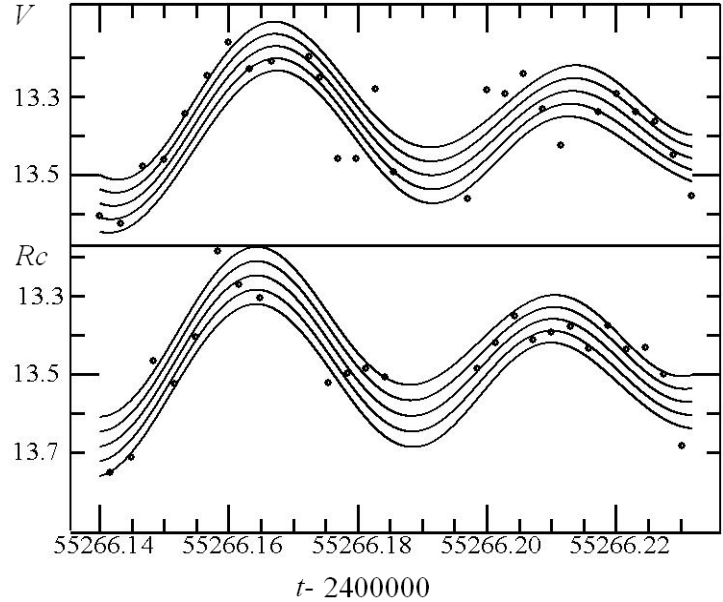

Fig. 2. The light curve of the EX Hya for one night (JD 2455266) of our observations in filters V and R with 2-periodic fit and ±1σ, ±2σ error corridors.

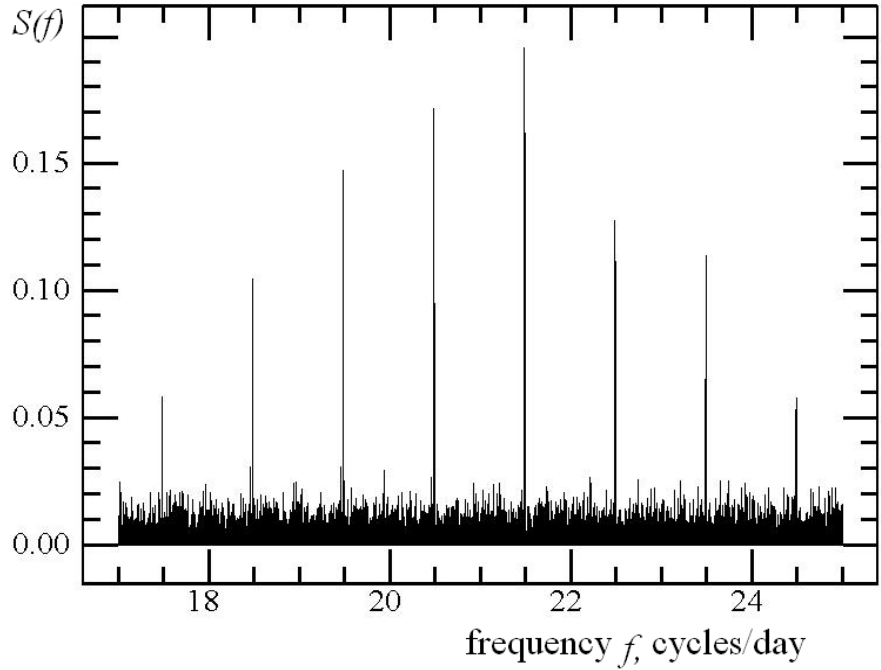

Fig. 3. Periodogram of EX Hya near the rotation period of the white dwarf. Maximum peak corresponds to a period of $0^d.046546$.

To determine the period for own observations, we conducted a periodogram analysis using a sine wave approximation [18,14]. Test-function $S(f)$, where the trial frequency $f=1/P$, is shown in Fig. 3. The highest peak corresponds to a period of $0^d.046546$. Other high peaks are shifted in frequency by $\pm k$ cycles per day, where $k$ is an integer.

Additionally, we analyzed observations from AAVSO (31908 observations JD 2434847-2455769), ASAS (688 observations JD 2451870-2455057), WASP (77 observations, JD 2453922-2454307) databases.

Periodograms on these data show low peaks, and the phase of the light curves do not show a good curve, which indirectly confirms the assumption about the variability of the period.

### Calculation of Phases in the Case of a Variable Period

In the simplest case of a constant period P, each moment of time $t$ corresponds to an integer cycle number $j$ and phase $\phi$, which can be calculated using the formula $E=j+\phi=(t-T_0)/P$, where $T_0$ is the initial epoch and $P$ is the period (e.g. [19]), i.e., $t=T_0+PE$. Although the period EX Hya are being studied for many decades, for obtaining the phase curves, are commonly used the values of the period and the initial epoch obtained for a short interval of observations. In the specific case of EX Hya, as well as the general case of the use of observations from the surveys, the calculation of phase for variable period curves and study phase curves for long term observations, becomes an actual topic. Therefore, we give the explicit form of the formula, which we used earlier in [12, 16] without a detailed description.

In the case of a variable period, one of the most common methods of research is the method of "O-C" diagrams that use only moments of typical events (minima of eclipsing stars, maxima of pulsating stars etc.). Moments are published in separate articles. The compiled database for hundreds of eclipsing stars is presented in a six-book monograph [20]. In recent years, the international databases of moments are available via the Internet (e.g., http://var2.astro.cz/EN/brno/index.php, www.aavso.org).

Parameters of variability are usually defined by modeling the deviations of moments of the characteristic events from a linear ephemeris, e.g.

$$O-C=T(E)-(T_{00}+P_0E)=Q_0+Q_1E+Q_2E^2+Q_3E^3+..+Q_nE^n+\underline{U}(E) \qquad (2)$$

Here $U(E)$ may be, e.g., periodic components associated with the presence of the third (and more) body in the system like stars or planets. The instantaneous value of the period

$$P(E)=dT(E)/dE=P_0+Q_1+2Q_2E+3Q_3E^2+..+nQ_nE^{n-1}+d\underline{U}(E)/dE \qquad (3)$$

This is slightly different from «differential» definitions $P_K(E)=(T(E+1)-T(E))$ by Kopal and Kurtz [21]. For differentiable functions $P(E)$, using the Lagrange formula, it is possible to get $P_K(E)=P(E+\delta(E))$. Here $0\leq\delta(E)\leq 1$ is the correction to the cycle number, which can be 0.5 in case of $dP/dE=2Q_2$=const. In general case of slow changes of the period, $|dP/dE|<<P$, $P(E)\approx P_K(E-0.5)$, and functions are almost the same. We use «instant» period for analysis.

Using the values of the coefficients of the equation (3), obtained when modeling the «special points», one can determine the value of $E$ for an arbitrary point in time. Theoretically, this means solution of the equation $T(E)=t$, however, in practice, it is possible to find the inverse function $E(t)$ analytically only for the simplest cases. So the preferred is numerical method of iterations:

$E_0=(t-T_0)/P_0$,

$$E_{k+1}=E_k - (t-T(E_k))/P(E_k) \qquad (4)$$

Obviously, if the right part is zero, only one initial approximation is necessary, in which we obtain the classical formula for a constant period. For polynomial models of degree higher than 2, we used this method in the study of the intermediate polar BG CMi [16] and others. In uniform (in respect to the cycle number $E$) period variation $dP/dE=2Q_2$,

$$t(E)=T_0+P_0E+Q_2E^2, \qquad (5)$$

and it's possible to get the inverse function $E(t)$ as the solution of a quadratic equation $T(E)=t$:

$$E = \frac{P_0}{2Q_2}\left(-1+\sqrt{1+\frac{4Q_2(t-T_0)}{P_0^2}}\right) = (-1+(1+4Q_2(t-T_0)/P_0^2)^{1/2})P_0/(2Q_2) = V(\varepsilon E_0)/\varepsilon \qquad (6)$$

where $\varepsilon=Q_2/P_0$, $V(x)=((1+4x)^{1/2}-1)/2$.

One can use the decomposition into the series

$$V(x)=x-x^2+2x^3-5x^4+14x^5-42x^6+132x^7-429x^8+1480x^9-4862x^{10}+\ldots \qquad (7)$$

Formally, the series converges for $|x|<0.25$, however, it is effective for $|x|<<1$, and really for $|x|<0.01$. From the formulae (6,7) it is possible to get:

$$E= E_0(1- x +2x^2-5x^3+14x^4-\ldots)= E_0- Q_2E_0^2/P_0+ 2Q_2^2E_0^3/P_0^2-\ldots \qquad (8)$$

**Analysis of variability of the rotation period of the white dwarf**

Previous analysis was made by Mauche et al. [23]. They published an ephemeris:

$$T(E)=2437699.8917(6)+0.046546484(9)E-7.3(4)\cdot 10^{-13}E^2+2.2(6)\cdot 10^{-19}E^3, \qquad (9)$$

and suggested the presence of a statistically significant coefficient $Q_3$. For the analysis of variability using the largest amount of data, we used either the moments of maxima determined from of our own and published patrol observations, or the published moments. In total, we have used 452 moment of maxima, which, due to limitations of the articles in volume, are published separately [24].

As the initial epoch, we have used the earliest timing in our list $T_{00}$= 2437699.8920 (Vogt et al. [25]). The initial value of the period $P_0=0.046546484^d$ (Mauche et al. [23]). Since the period variation is significant, the sequence of cycles should be corrected by one or even two cycles, which is indicated in the appropriate column of the Table. 1 [24]. Using the program MCV [12], we determined the statistically optimal degree of the polynomial for the O-C approximation. It is equal to two. Parameter $Q_3$ is not statistically significant, i.e. considerably more data do not confirm the assumption by Mauche et al. [23]. Using specified data, we obtained the ephemeris

$$T_{max}=2437699.89079(59)+0.0465464808(69)\cdot E-6.3(2)\cdot 10^{-13}E^2. \qquad (10)$$

Thus, on the specified time interval of 49 years, the deviations from the model of uniform period change are not statistically significant.

O-C diagram and it's ±1σ, ±2σ error corridors are shown on Fig. 4. Note that the errors of smoothing curve were calculated using correct formulae taking into account the covariance matrix of the errors of the coefficients rather than reduced, taking into account only the errors of the coefficients (see [18,15]).

Using the coefficients of the resulting relation (10), we calculated the phases using the formulae (6,8). The obtained phase curves according to the ASAS, TzecMaun V and R and AAVSO data are shown in Fig. 5. The first 3 curves correspond to the minima $m \sim 13.5^m$, and show almost sinusoidal shape. Significant scatter on the phase curve is due to not only errors of observation, but also to a significant orbital variability, comparable in amplitude with the spin one. Some AAVSO observations show outbursts up to $9.3^m$, however, a series of observations during the night correspond to the level of $10.5m$.

The amplitude of the sine changes in the bright state (at an outburst) is the same (about $0.3^m$), as in a low state. This is consistent with the assumption that the brightening of the accretion disk and accretion column occurs approximately in proportion.

The significant shift of the maximum phase by phase-0.2 during the outburst is to be noted. A similar effect of the negative phase shift with increase of brightness is observed in the eclipsing polar OT J071126.0+440405 = CSS 081231:071126+440405 [26,27] and is explained by the smaller distance, at which plasma is captured by the magnetic field, when the mass flow increases.

Two light curves obtained during the outburst, slightly differ, as seen in Fig. 5. Unfortunately, because of such differences, a detailed comparison of the characteristics of the light curves at minimum and maximum is impossible, and in the future, the observations during a rare outbursts are required. Theoretical models of outbursts are discussed by Mhahlo et al [28].

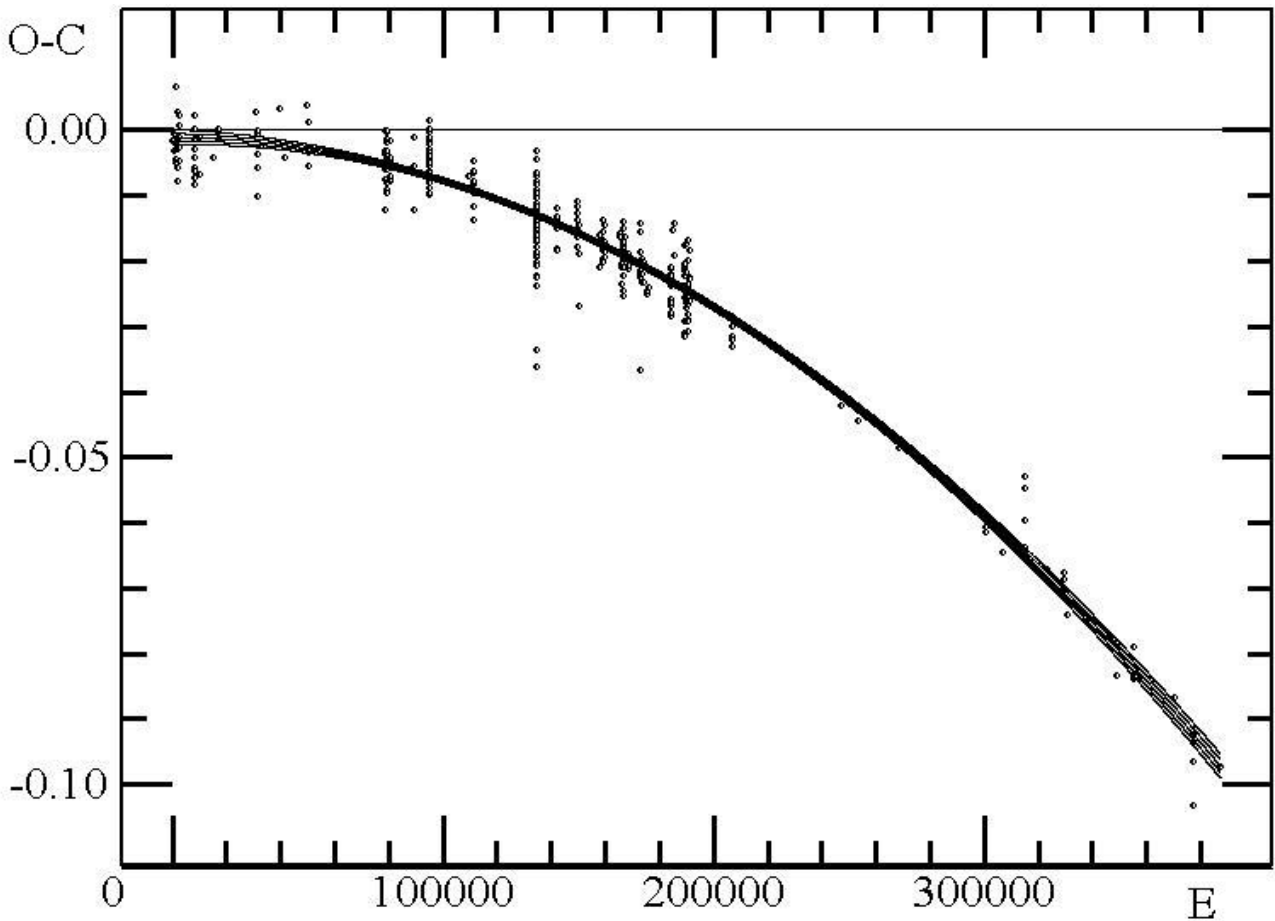

Fig. 4. O-C diagram for the spin maxima of EX Hya, calculated for the values of the initial epoch $T_{00}$= 2437699.8920 (Vogt et al. [25]) and the period $P_0$= 0.046546484$^d$ (Mauche et al. [23]) according to the Table 1 [24].

**Discussion**

EX Hya belongs to ourbursting intermediate polars, and the periodicity associated with the rotation of the white dwarf, is well expressed. This differs from DO Dra, often mistakenly called YY Dra, where strict periodicity is absent. However, sometimes there are «transient periodic oscillations» [4], a period which varies from night to night, and, probably, connected with interaction of plasma clouds on the inner part of the accretion disk with the magnetic field of a rapidly rotating white dwarf.

Thus, we can conclude that the accretion columns above the atmosphere of the white dwarf in EX Hya are much more powerful (relatively to other sources of radiation), than DO Dra. Due to this, EX Hya should be classified as an "outbursting intermediate polar", and DO Dra, as a "magnetic dwarf nova". Despite the existence of outbursts in both systems, which differ from outbursts from the «non-magnetic» U Gem type dwarf novae stars, a distinction between systems is present, and the subtype classification should be used.

The characteristic time of the change of the period (acceleration of rotation)

$$\tau = P_0/|dP/dt| = P_0^2/|2Q_2| = 4.67(14) \cdot 10^6 \text{ years} \qquad (11)$$

This is significantly more than the synchronization time $\tau_s = 96.7 \pm 1.5$ and time of period changes $\tau = 3.5(1) \cdot 10^4$ for V1432 Aql [29] and in 5 times larger than $\tau$ for BG CMi [16]. It should be noted that some systems for decades does not show a statistically significant change of the period (e.g. 1RXS J180340.0+401214 = RXJ 1803 = V1323 Her), and unsolvable riddle about the failure of the phase fluctuations (on data from other author) for 0.5 period was connected with the fact that the authors of one of the published articles did not consider that the «maximum brightness» corresponds to the «minimum magnitude» [30].

Other intermediate polars show for decades either an increase of the period, or its subsequent decrease (e.g. FO Aqr [31], V405 Aur [32]).

According to current theoretical models (e.g. [33]), the change of the equilibrium spin period occurs at essentially higher characteristic time of billions years. Thus, the observed acceleration of the rotation is a short-term phenomenon. If it was associated with the suddenly increased mass flow, it should be associated with an increased flux of radiation. The light curve has outbursts, however, they are numerous, and, due to the inertia of the white dwarf, a temporary increase in flow during outbursts leads to a transient increase in luminosity, but the long-term period change is affected by the average level of accretion only.

Change of the equilibrium period with characteristic time of hundreds of years can be attributed to fluctuations in the accretion rate due to the magnetic activity of the secondary component - red dwarf (Bianchini [34], Andronov and Shakun [35]) or by small changes in the distance between components due to gravity of low-mass third body (e.g. red or brown dwarf) (Andronov and Chinarova [36]). Another mechanism of change of the equilibrium period, which can take place at constant accretion rate, is the precession of the rotation axis of the magnetic white dwarf([37], [38], [39]).

On the website http://www.ukaff.ac.uk/movies.shtml, there is animated model of the system EX Hya. Accretion disk is absent, the flow is deflected by the magnetic field and subsequently falls alternately on the magnetic poles.

Moment of force, leading to the observed acceleration is equal to $dJ/dt = I d\omega/dt$.
One can estimate the angular momentum using the approximating formula

$$I = (4.095 - 2.795 M - 3.207 \cdot \exp(-2.455 M)) \cdot 10^{50} \text{ g} \cdot \text{cm}^2$$

[40], where M is the mass of the white dwarf in the solar masses. The mass estimate $M = 0.42\ M_\odot$ [41] corresponds to $I = 1.777 \cdot 10^{50}$ g·cm$^2$ and $dJ/dt = 1.87 \cdot 10^{33}$ g·cm$^2$/s. Estimate of the accretion rate is $dM/dt = 1.36 \cdot 10^{15}$ g/s, assuming the evolution of the binary due to gravitational radiation [33]. Estimate of the corresponding radius of the capture of the accretion flow $Rc = 3.4 \cdot 10^{10}$ cm exceeds the value of the distance between the components of $a = 3.0 \cdot 10^{10}$ cm. Thus, it is necessary to

determine the system parameters more precisely for a self-consistent model. Alternative mass definitions presented by Beuermann and Reinsh [42]: $M_2$=0.108±0.008 $M_\odot$, $M_1$=0.790±0.026 $M_\odot$. In this case, according to Kepler's third law, $a$=(3.66±0.004)·10$^{10}$cm and corresponding parameters $I$=1.426·10$^{50}$ g·cm$^2$, d$J$/d$t$=1.50·10$^{33}$ g·cm$^2$/s, $R$c=1.16·10$^{10}$см=0.318$a$. It is close to the radius of the Roche lobe in the direction perpendicular to the line of centers for the given values of the mass ratio. Thus, the obtained value of the acceleration of the rotation of the white dwarf is in a better agreement with parameters [42] than [41].

The O-C diagram based on decades of observations is well described by the quadratic parabola, and there are no grounds to assume that the acceleration rate of rotation of the white dwarf is slowing, and the possibility of cyclic fluctuations. In the absence of acceleration deceleration, the more likely suggestion is that the rotation period of the white dwarf will decrease significantly to the «main group» at the diagram "$P_{spin}$-$P_{orb}$ in the wide area of $P_{spin}$~0.1$P_{orb}$ [1]. Norton et al. [33] separated the group of stars of EX Hya type, as that corresponding to a weak magnetic field of the secondary star, which fills its Roche lobe.

To choose between these models, it is necessary to continue regular observations of EX Hya.

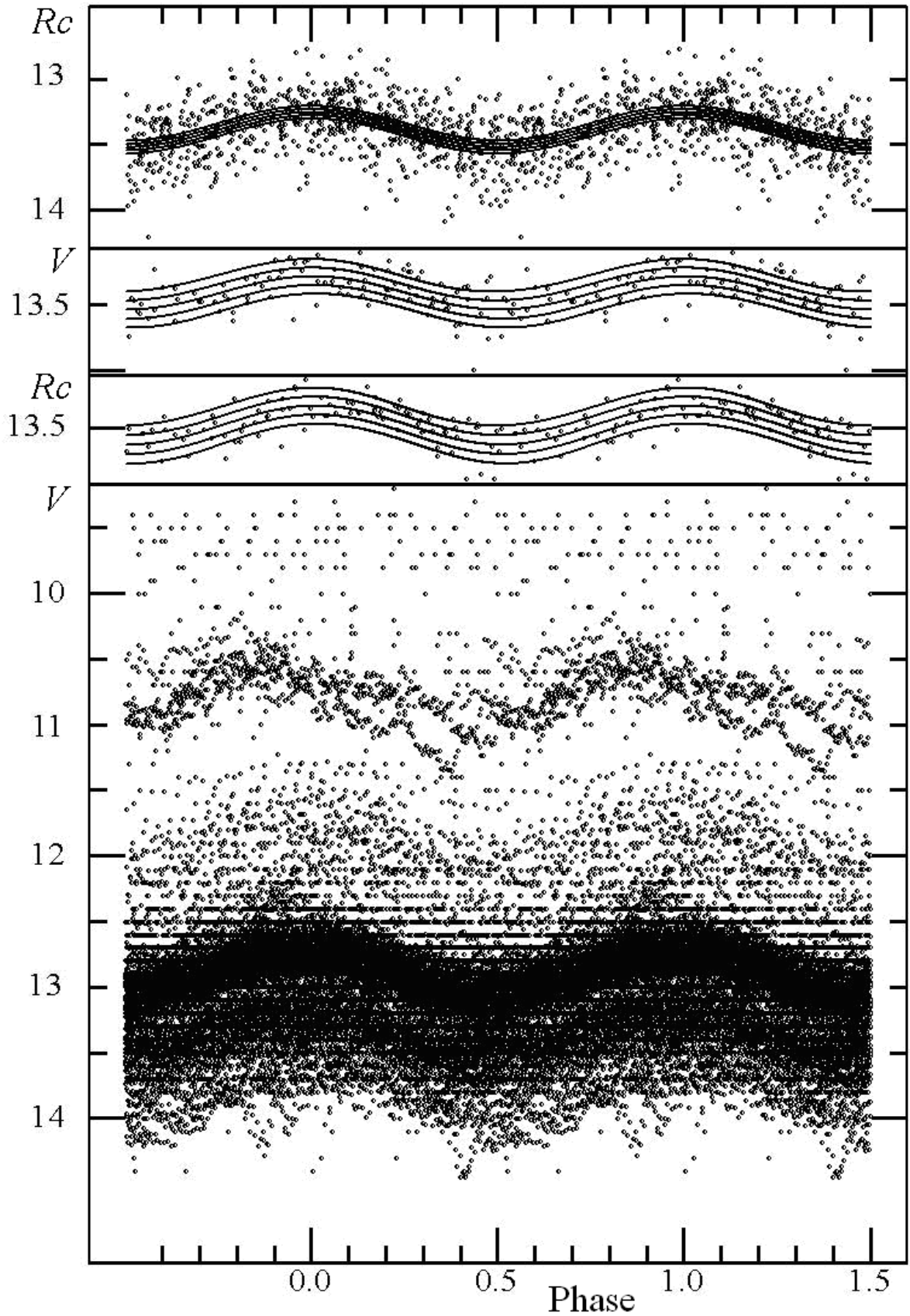

Fig. 5. The phase curves EX Hya, calculated for a parabolic ephemerides (10) according to the data from ASAS (1), TzecMaun V (2), TzecMaun R (3), AAVSO (4). Figures correspond to the number of the pattern above.

*Acknowledgements:* Our observations obtained using the remote access to the Tzec Maun Observatory. We used observations from the databases of projects AAVSO, ASAS, SuperWASP. The authors thank to V. P. Grinin and A.V. Baklanov for useful discussion.